# On the quantification of the dissolved hydroxyl radicals in the plasma-liquid system using the molecular probe method

Yupengxue Ma[1], Xinning Gong[1], Bangbang He[1], Xiaofei Li[2], Dianyu Cao[2], Junshuai Li[3], Qing Xiong[4], Qiang Chen[1,*] Bing Hui Chen[2], and Qing Huo Liu[5]

[1]Fujian Provincial Key Laboratory of Plasma and Magnetic Resonance, Institute of Electromagnetics and Acoustics, Department of Electronic Science, Xiamen University, Xiamen 361005, China

[2]Department of Chemical and Biochemical Engineering, National Engineering Laboratory for Green Productions of Alcohols-Ethers-Esters, College of Chemistry and Chemical Engineering, Xiamen University, Xiamen 361005, China

[3]School of Physical Science and Technology, Lanzhou University, Lanzhou 730000, China

[4]State Key Laboratory of Power Transmission Equipment & System Security and New Technology, Chongqing University, Chongqing 400044, China

[5]Department of Electrical and Computer Engineering, Duke University, Durham, NC 27708, USA

*Corresponding author: chenqiang@xmu.edu.cn

**ABSTRACT:**

Hydroxyl (OH) radical is the most important reactive species produced by the plasma-liquid interactions, and the OH in the liquid phase (dissolved OH radical, $OH_{dis}$) takes effect in many plasma-based applications due to its high reactivity. Therefore, the quantification of the $OH_{dis}$ in the plasma-liquid system is of great importance, and a molecular probe method usually used for the $OH_{dis}$ detection might be applied. Herein we investigate the validity of using the molecular probe method to estimate the [$OH_{dis}$] in the plasma-liquid system. Dimethyl sulfoxide is used as the molecular probe to estimate the [$OH_{dis}$] in an air plasma-liquid system, and the partial $OH_{dis}$ is related to the formed formaldehyde (HCHO) which is the $OH_{dis}$-induced derivative. The analysis indicates that the true concentration of the $OH_{dis}$ should be estimated from the sum of three terms: the formed HCHO, the existing OH scavengers, and the $OH_{dis}$ generated $H_2O_2$. The results show that the measured [HCHO] needs to be corrected since the HCHO destruction is not negligible in the plasma-liquid system. We conclude from the results and the analysis that the molecular probe method generally underestimates the [$OH_{dis}$] in the plasma-liquid system. If one wants to obtain the true concentration of the $OH_{dis}$ in the plasma-liquid system, one needs to know the destruction behavior of the $OH_{dis}$-induced derivatives, the information of the OH scavengers (such as hydrated electron, atomic hydrogen besides the molecular probe), and also the knowledge of the $OH_{dis}$ generated $H_2O_2$.

Discharge plasma consists of approximately equal number of energetic ions and electrons, and these plasma species can cause a great number of physical and chemical processes when plasma is in contact with a liquid. By exploiting these processes and their derivative highly reactive species, the plasma-liquid system can find applications in many fields such as water treatment [1-3], plasma medicine [4, 5], nanomaterials synthesis [6-18], and food processing [19-21]. Thereby, it is of great importance to identify and quantify the reactive species in the plasma-liquid interactions, especially for water treatment and plasma medicine in which the treatment efficiency is largely related to the dissolved or transported reactive species. One of the most important reactive species generated by the plasma-liquid interactions is the hydroxyl (OH) radical. Because of its high reactivity (reduction potential of ~2.8 eV), OH radical has a short lifetime and can only reach a very thin layer of the bulk liquid. However, many applications of the plasma-liquid system are mainly or partially based on the dissolved OH radical-induced processes, such as water treatment and plasma medicine. Thus, it is of great significance to quantify the dissolved OH ($OH_{dis}$) radicals generated in the plasma-liquid system.

Previously, the $OH_{dis}$ generated from techniques such as Fenton reaction, UV light irradiation of $NO_3^-$ or $NO_2^-$ in solution rather than the plasma-liquid system has been quantified by several methods [22]. As OH radicals are highly reactive and have a limited diffusion length in liquid, a direct detection is rather difficult. Therefore, indirect methods have been developed, and one of them is the molecular probe (MP) method. The strategy is to form a long-lived OH-induced derivative by using a molecule to trap or react with OH radicals at first, and then the $OH_{dis}$ is indirectly detected by probing the OH-induced derivative. These derivative detection methods can be performed by electron spin resonance (ESR) spectroscopy [23], high performance liquid chromatography (HPLC) [24], and fluoroluminescence (FL) spectroscopy [25]. In order to obtain correct results, it is worth noting that the used MP should react with the OH with a fast speed and the derivatives must be stable under the attack of highly reactive OH radicals.

Several works [26-31] have been reported on applying the MP method to detect the $OH_{dis}$ in the plasma-liquid system. However, it is well known that the plasma-liquid system is very complex, and the reactive species generated by the plasma-liquid interactions consist of not only OH radicals, but also energetic ions, hydrated electrons, and atomic hydrogen etc. In these complicated conditions, there might exist other OH consumed sources, and the OH-induced derivatives might not be as stable as in the sole OH environment. Thereby, we must reconsider the validity of the MP method on the $OH_{dis}$ detection in the plasma-liquid system.

When the OH radicals produced by the plasma-liquid interactions enter a liquid containing a MP, they will be consumed quickly within a thin liquid layer by a series of parallel reactions mainly with the MP, existing potential OH scavengers ($S_i$), and OH itself to produce related products of $P_{MP}$, $P_i$, and $H_2O_2$

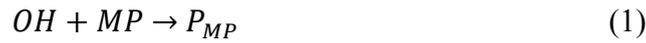

$$OH + MP \rightarrow P_{MP} \qquad (1)$$

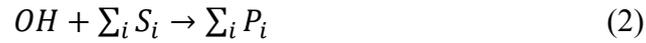

$$OH + \sum_i S_i \rightarrow \sum_i P_i \qquad (2)$$

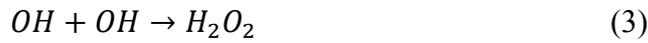

$$OH + OH \rightarrow H_2O_2 \qquad (3)$$

The change of OH in solution with respect to time (d[OH]/dt) can be expressed by

$$\frac{d[OH]}{dt} = G_{OH} - k_{MP}[MP]^a[OH]^b - \sum_i k_{Si}[S_i]^{x_i}[OH]^{y_i} - k_{H2O2}[OH]^c \quad (4).$$

where $G_{OH}$ is the rate of OH radicals dissolved in the liquid, $K_{MP}$, $K_{Si}$, and $K_{H2O2}$ are the rate constants of OH with MP, the $i$th OH scavenger, and OH itself (to form $H_2O_2$), respectively. [MP], [OH], and [$S_i$] are the concentrations of MP, OH, the $i$th scavenger in liquid, respectively. a, b, $x_i$, $y_i$, c are the partial orders of related reactions. In a pseudo steady state (d[OH]/dt=0), the $G_{OH}$ can be expressed as

$$G_{OH} = k_{MP}[MP]^a[OH]^b + \sum_i k_{Si}[S_i]^{x_i}[OH]^{y_i} + k_{H2O2}[OH]^c \qquad (5).$$

If the last two terms in Eq. 5 are negligible, and $P_{MP}$ is also stable under the plasma treatment, the total concentration of dissolved OH ($[OH_{dis}]$) during the plasma-liquid interactions can be obtained by

$$[OH_{dis}] = \int G_{OH} dt = \int k_{MP}[MP]^a[OH]^b dt \qquad (6).$$

These assumptions are used in most of the cases for detection of the $OH_{dis}$ in the plasma-liquid system [26-31]. But these assumptions might not be valid since there exists various highly reactive species in the plasma-liquid system. In order to verify these assumptions, we design an experiment to inspect the $[OH_{dis}]$ in a plasma-liquid system using the MP method. There exist many MPs for the $OH_{dis}$ detection for the plasma-liquid system, such as 5,5-dimethyl-1-pyrrolineN-oxide [31], disodium salt of terephthalic acid [26, 32], terephthalic acid [27, 33, 34], salicylic acid [35, 36], and dimethyl sulfoxide (DMSO) [26, 32, 37]. To focus our discussion, we choose one of them, DMSO, which is miscible with water. The DMSO can react with OH to form OH-induced derivative, formaldehyde (HCHO).

The experimental setup is illustrated in Fig. 1. The cylinder-like plasma reactor is made from polytetrafluoroethylene (detail parameters of the reactor can be found in Ref. [38]). The plasma treated liquid (200 ml) is circulated by a peristaltic pump with a flow rate of 100 ml/min. The flowing silicone tube is 3 mm and 5 mm in inner and outside diameters, respectively. To generate the atmospheric pressure discharge plasma in open air, a direct current power source (BOHER HV, LAS-20 KV-50 mA, positive polarity) is applied to a solid tungsten steel electrode (4 mm in diameter), and the plasma is generated between the solid electrode and the flowing liquid surface (discharge gap of 3 mm). A graphite rod (5 mm in diameter) is grounded at the bottom of the solution to act as an inert electrode. A 50-k$\Omega$ resistor is connected in series with the tungsten steel electrode to avoid the plasma transfer from glow-like discharge to arc. The discharge current (fixed for 30 mA in this study) is achieved from dividing the voltage across a 10-$\Omega$ resistor which is in series connected with the graphite electrode. After the plasma treatment, the treated liquid is mixed in a blending beaker by a magnetic stirrer. Oxygen gas (150 sccm, 99.9% in purity) is bubbled into the blending beaker for the oxygen-involved reactions mentioned later. The liquid is an aqueous solution of DMSO or a mixture of DMSO and HCHO.

DMSO (>99.8%), HCHO (37%), and acetic acid (≥99%) were purchased from Xilong Scientific Co., Ltd. Ammonium acetate (≥99%) and acetylacetone (≥99.3%) were purchased from Sinopharm Chemical Reagent Co., Ltd.

Hantzsch reaction [39] was used for the colorimetric estimation of HCHO concentration. We prepared aqueous solutions which contain ammonium acetate (50 g), acetic acid (6 ml), and acetylacetone (0.5 ml) in 100 ml purified water. Samples (1 ml) of the plasma treated liquid were taken from the beaker in a fixed interval, and then they were mixed with 1 ml the prepared solution. The mixed solutions were bathed in heated water (60 $^{o}$C) for 15 min, and then cooled to room temperature. Absorbances were measured for the mixed solutions in a quartz cell (10 mm in optical thickness) by combining an optical emission spectroscopy (Ocean Optics USB 2000+) and a tungsten halogen light source (Ocean Optics HL-2000). It is well known that the absorbance intensity at 410 nm of the mixed solution is proportional to the HCHO concentration [39]. The linear relationship between the HCHO concentration and the absorbance intensity at 410 nm was achieved by measuring a series of standard HCHO solutions.

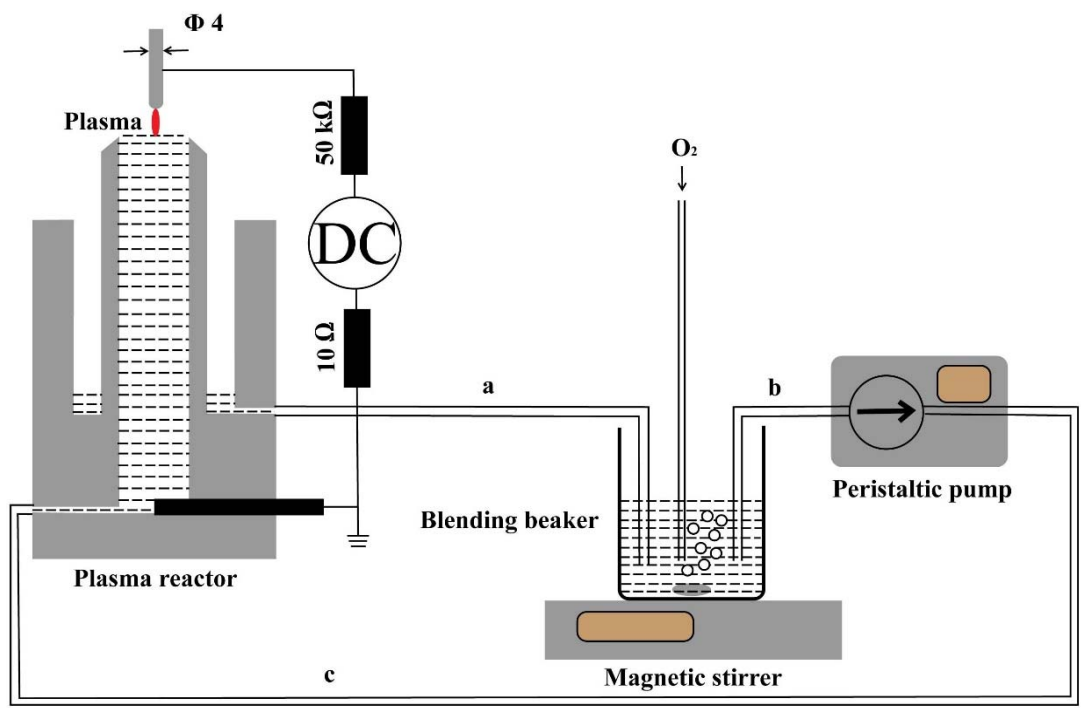

**Figure 1.** Schematic diagram of the experimental setup. Lengths of silicone tubes a, b, c are 20 cm, 20 cm, and 25 cm, respectively. The inner and outside diameters of the tubes are 3 and 5 mm, respectively. Insider the pump, there is a 10 cm-silicone tube (4 and 6 mm in inner and outside diameters) to connect tubes b and c.

DMSO can react with the $OH_{dis}$ at a very fast rate constant of $6.6 \times 10^9$ $M^{-1}s^{-1}$ [40], producing methanesulfinic acid ($CH_3SOOH$), methyl radical ($CH_3$), and at the present of oxygen, HCHO and methanol ($CH_3OH$) via Eqs. 7-9,

$$(CH_3)_2SO + OH \rightarrow CH_3SOOH + CH_3 \qquad (7)$$

$$CH_3 + O_2 \rightarrow CH_3OO \qquad (8)$$

$$2CH_3OO \rightarrow HCHO + CH_3OH + O_2 \qquad (9).$$

The produced methanesulfinic acid [41, 42] or formaldehyde [24, 26] has been taken as the OH$_{dis}$-induced derivative for the OH$_{dis}$ detection. The [OH$_{dis}$] is usually considered to stoichiometrically relate to the measured concentrations of methanesulfinic acid (1:1) or formaldehyde (2:1).

In fact, methanol produced by Eq. 9 can react with the OH$_{dis}$ at a high rate constant of $9.7 \times 10^8$ M$^{-1}$s$^{-1}$ [40], and also produce HCHO at the presence of oxygen via Eqs. 10-11 [43],

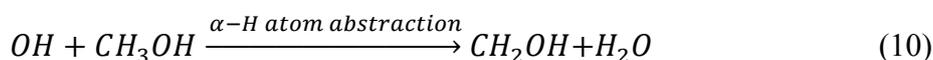

$$OH + CH_3OH \xrightarrow{\alpha-H \; atom \; abstraction} CH_2OH + H_2O \qquad (10)$$

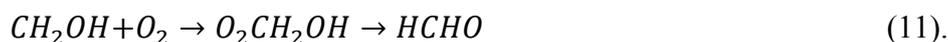

$$CH_2OH + O_2 \rightarrow O_2CH_2OH \rightarrow HCHO \qquad (11).$$

Therefore, the produced HCHO might come from two pathways by a series of reactions,

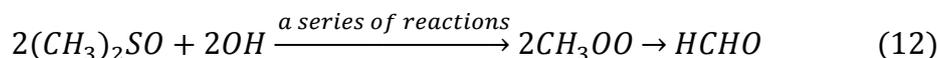

$$2(CH_3)_2SO + 2OH \xrightarrow{a \; series \; of \; reactions} 2CH_3OO \rightarrow HCHO \qquad (12)$$

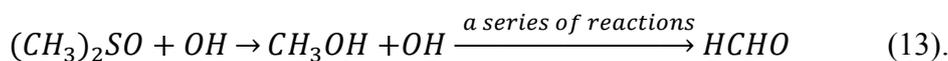

$$(CH_3)_2SO + OH \rightarrow CH_3OH + OH \xrightarrow{a \; series \; of \; reactions} HCHO \qquad (13).$$

Based on Eqs. 7-9, one can find that two moles of OH$_{dis}$ produce one mole of HCHO and two moles of CH$_3$OH. If Z ($0 \leq Z \leq 1$) mole of the CH$_3$OH produced from two moles of OH$_{dis}$ takes part in Reaction 13, we have the following relation: 2+Z moles of OH$_{dis}$ radicals correspond to 1+Z moles of HCHO. That is to say, the OH$_{dis}$ (used to produce HCHO) concentration is related to the measured concentration of HCHO by a ratio of (2+Z) : (1+Z) rather than 2 : 1.

Based on Eqs. 7-13, the d[HCHO]/dt and d[CH$_3$OH]/dt can be expressed as

$$\frac{d[HCHO]}{dt} = 2k_{DMSO}[DMSO]^a[OH]^b + k_{CH_3OH}[Z \cdot CH_3OH]^c[OH]^d$$
$$- \sum_i k_{HCHOi}[HCHO]^{e_i}[X_i]^{f_i} \qquad (14)$$

$$\frac{d[CH_3OH]}{dt} = 2k_{DMSO}[DMSO]^a[OH]^b - \sum_i k_{CH_3OH}[CH_3OH]^{g_i}[Y_i]^{h_i} \qquad (15).$$

where $X_i$ and $Y_i$ are the $i$th reactants in the liquid able to destruct HCHO and $CH_3OH$, respectively. For instance, the $OH_{dis}$ can react with HCHO to form formate and the final product carbon dioxide in the presence of $O_2$ [44]. $e_i$, $f_i$, $g_i$, $h_i$, are the partial orders of related reactions.

In this experiment, only a small ratio of the total DMSO is consumed, [DMSO] can be considered to be constant during the plasma treatment. In addition, for a pseudo steady state ($d[OH]/dt=0$), [OH] is a constant. Consequently, $2K_{DMSO}[DMSO]^a[OH]^b$ can be represented by a constant of $k_1$. We found the third term in Eq. 14 is a constant during the plasma treatment by measuring the [HCHO] variation as shown later in Fig. 3. That means the plasma-induced destruction of HCHO is a pseudo zero-order reaction (reaction rate of $k_2$). If we further assume that plasma-induced destruction of $CH_3OH$ is also a pseudo zero-order reaction (reaction rate of $k_3$). Therefore, we have

$$\frac{d[CH_3OH]}{dt} = k_1 - k_3 \qquad (16)$$

$$[CH_3OH] = (k_1 - k_3)t, \quad ([CH_3OH] = 0 \ at \ t = 0) \qquad (17).$$

If we assume c=d=1, then

$$\frac{d[HCHO]}{dt} = k_1 + Z \cdot [(k_1 - k_3)t] - k_2 = Z \cdot (k_1 - k_3)t + (k_1 - k_2) \qquad (18)$$

$$[HCHO] = Z \cdot (k_1 - k_3)t^2 + (k_1 - k_2)t + C_1 \qquad (19).$$

From Eq. 19, the true concentration of $OH_{dis}$-induced HCHO ($[HCHO]_{true}$) should be

$$[HCHO]_{true} = [HCHO] + k_2t \qquad (20).$$

Figure 2 presents the measured [HCHO] for the plasma-treated DMSO solutions (10 mM) with different initial [HCHO]. The changes of [HCHO] ($\Delta$[HCHO]) are summarized in Table 1. The temporal [HCHO] can be fitted well by a two-order polynomial as shown in Fig 2. The coefficients of the polynomials are given in Table 2. Therefore, Z in Eq. 19 is not zero. Evidently, $\Delta$[HCHO] depends on the initial [HCHO], and its absolute value decreases with increasing [HCHO]. Simply, there must exist the HCHO destruction and the generation of HCHO from $CH_3OH$ during the plasma treatment.

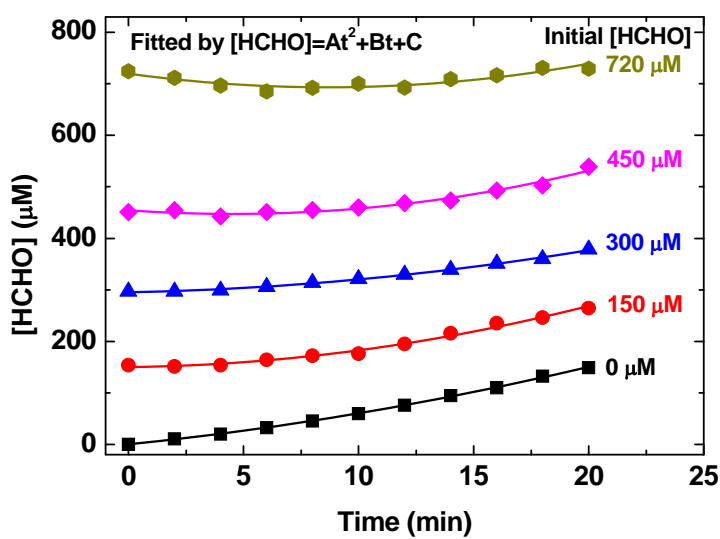

**Figure 2.** Temporal [HCHO] for plasma treatment of aqueous DMSO (10 mM) with different initial concentrations of HCHO. The continuous lines are the data fitted with two-order polynomials, [HCHO]=$At^2$+Bt+C.

**Table 1.** [HCHO] changes in DMSO solutions (10 mM) after 20-min plasma treatment.

| Initial [HCHO] ($\mu$M) | Final [HCHO] ($\mu$M) | $\Delta$[HCHO] ($\mu$M) |
|---|---|---|
| 0 | 150 | 150 |
| 150 | 260 | 110 |
| 300 | 380 | 80 |
| 450 | 540 | 90 |
| 720 | 730 | 10 |

**Table 2.** Coefficients of the two-order polynomials by fitting [HCHO] with $At^2+Bt+C$, and the reaction rate $k_1$ ($2K_{DMSO}[DMSO]^a[OH]^b$) calculated from $B=k_1-k_2$.

| Initial [HCHO] ($\mu$M) | A ($\mu$M min$^{-2}$) | B ($\mu$M min$^{-1}$) | $k_2$ ($\mu$M min$^{-1}$) | $k_1$ ($\mu$M min$^{-1}$) |
|---|---|---|---|---|
| 0 | 0.15 | 4.55 | 0 or 9.78 | 4.55 or 14.33 |
| 150 | 0.27 | 0.56 | 9.78 | 10.34 |
| 300 | 0.16 | 0.94 | 9.78 | 10.72 |
| 450 | 0.35 | -3.21 | 9.78 | 6.57 |
| 720 | 0.36 | -6.15 | 9.78 | 3.63 |

In order to investigate the HCHO destruction by the plasma-liquid interactions, a HCHO solution with an initial [HCHO] of 628 $\mu$M was treated by plasma for 2 h. Figure 3 shows the relationships between the [HCHO] and the plasma treatment time. We can find that the destruction of HCHO by plasma is roughly linear for [HCHO] is higher than 20 $\mu$M, and the destruction rate is estimated to be -9.78 $\mu$M/min. V.V. Kovačević1 et al. found that the HCHO with a concentration of 50 $\mu$M is stable in water when a dielectric barrier discharge in direct contact with water [37], while in our case, the HCHO is stable under the plasma treatment as [HCHO]$\leq$20 $\mu$M. These results imply that plasma-induced destruction of HCHO is a pseudo zero-order reaction, i.e., the third term of Eq. 14 is a constant ($k_2$).

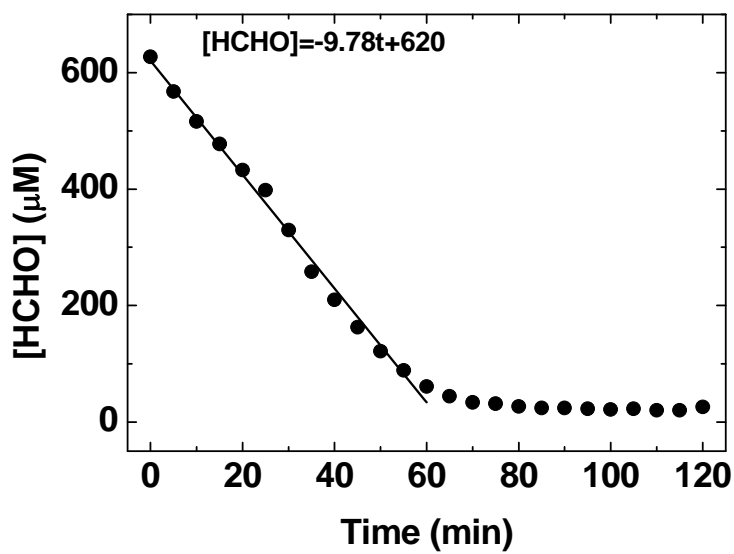

**Figure 3.** Temporal [HCHO] for plasma treated aqueous HCHO solution with an initial [HCHO] of 628 μM.

If we use $k_2=0$ μM/min for the solutions with [HCHO]≦20 μM, and $k_2=9.78$ μM/min for the solutions with [HCHO]>20 μM, we can correct the [HCHO] in Fig. 2 to the [HCHO]$_{true}$ in Fig. 4. The changes of [HCHO] $_{true}$ (Δ[HCHO]$_{true}$) are summarized in Table 3.

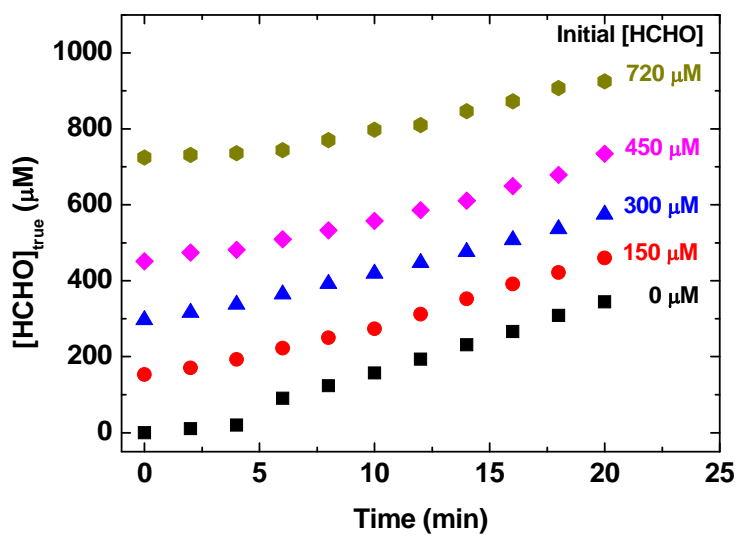

**Figure 4.** Temporal concentrations of the HCHO (containing the plasma-destructed HCHO) ([HCHO]$_{true}$) for plasma treated aqueous DMSO (10 mM) with different initial [HCHO].

**Table 3.** Changes of [HCHO]$_{true}$ in DMSO solutions (10 mM) after 20-min plasma treatment.

| Initial [HCHO] ($\mu$M) | Final [HCHO]$_{true}$ ($\mu$M) | $\Delta$[HCHO]$_{true}$ ($\mu$M) |
|---|---|---|
| 0 | 344 | 344 |
| 150 | 460 | 310 |
| 300 | 574 | 274 |
| 450 | 735 | 285 |
| 720 | 925 | 205 |

If we only consider the OH$_{dis}$ related to the OH$_{dis}$-generated HCHO, we obtain

$$[OH]_{HCHO} = \frac{2+Z}{1+Z}[HCHO]_{true} \qquad (21).$$

In fact, considering Eq. 5 and the fact of the existence of OH scavengers (S$_i$) in the plasma-liquid system, the OH$_{dis}$ is consumed not only by forming HCHO, but also by reacting with the existing OH scavengers such as hydrated electrons, atomic hydrogens [40] as well as OH itself, and thus [OH$_{dis}$] in our case can be expressed as

$$[OH_{dis}] = \int G_{OH} dt$$
$$= \int \left( k_{DMSO}[DMSO]^a [OH]^b + \sum_i k_{Si}[S_i]^{x_i}[OH]^{y_i} \right.$$
$$\left. + k_{H2O2}[OH]^c \right) dt \qquad (22).$$

In Eq. 22, besides the first term which can be related to the [HCHO]$_{true}$, there exist OH$_{dis}$ consumed by the OH scavengers such as hydrated electron, atomic hydrogen [40] as well as HCHO (the second term), and the transfer of OH to H$_2$O$_2$ (the third term).

We can calculated the concentration of the HCHO related $OH_{dis}$ by Eq. 21, if we know the value of Z. The second term is difficult to be obtained due to the lack of information for the existing $OH_{dis}$ scavengers in the plasma-liquid system. Because the $H_2O_2$ in liquid generated by the plasma-liquid interactions are from two sources: one is from the gaseous OH combination, and the other is from the combination of dissolved OH radicals, we are not able to quantify the $[OH_{dis}]$ by measuring the $[H_2O_2]$ in the plasma-treated liquid. Consequently, only the concentration of HCHO related $OH_{dis}$ can be estimated by this method if the value of Z is known, and the MP method actually underestimate the concentration of $OH_{dis}$ produced by the plasma-liquid interactions.

To summarize, DMSO was used as the molecular probe to estimate the concentration of $OH_{dis}$ generated by the plasma-liquid interactions. The results and the analysis imply that the concentration of the $OH_{dis}$-induced derivative, HCHO (used to estimate the $[OH_{dis}]$), needs to be corrected by considering the HCHO destruction and the HCHO production from the $OH_{dis}$ produced methanol. From the corrected [HCHO], we can only estimate the concentration of $OH_{dis}$ related to the HCHO, while the true $[OH_{dis}]$ is in fact difficult to be evaluated due to the ignorance of the information of existing potential $OH_{dis}$ scavengers and the knowledge of the $OH_{dis}$-generated $H_2O_2$. Generally, when using other molecular probes such as terephthalic acid and salicylic acid, one will still encounter the aforementioned difficulties. The estimation of $[OH_{dis}]$ by the molecular probe in the plasma-liquid system is usually underestimated due to the complicated reaction environment. One can estimate the $[OH_{dis}]$ related to the $OH_{dis}$-induced derivatives (used to probe the $OH_{dis}$), but the concentration of $OH_{dis}$- induced derivatives might need a correction by considering the destruction of derivatives. Moreover, it is worth noting that the destruction of the $OH_{dis}$-induced derivatives is system dependent since the highly reactive species are system dependent. Thereby, one needs to measure the destruction of the $OH_{dis}$-induced derivatives each time for different plasma-liquid systems.

## Acknowledgments

The work was partially supported by National Natural Science Foundation of China (Grant Nos.: 11405144, 11304132, and 61376068), and the Fundamental Research Funds for the Central Universities (Grant No: 20720150022). One of the XQ thanks the financial supports from the Graduate Scientific Research and Innovation Foundation of Chongqing (No. CYS17007), and National "111" Project of China (Grant No: B08036).